\documentclass[12pt]{article}
\usepackage{amsfonts,amsmath}
\usepackage{hyperref}

\makeatletter \@addtoreset{equation}{section} \makeatother

\newcommand{\be}{\begin{equation}}
\newcommand{\ee}{\end{equation}}
\newcommand{\bee}{\begin{eqnarray}}
\newcommand{\beee}{\begin{array}}
\newcommand{\eee}{\end{eqnarray}}
\newcommand{\eeee}{\end{array}}

\newcommand{\ga}{\alpha}
\newcommand{\pa}{{\dot{\ga}}}
\newcommand{\pb}{{\dot{\gb}}}

\newcommand{\gb}{\beta}
\newcommand{\gga}{\gamma}

\newcommand{\W}{{\cal W}}

\newcommand{\Ll}{{\cal L}}

\newcommand{\ie}{{\it i.e.,} }

\def\ck{{\mathcal K}}

\newcommand{\gvep}{\varepsilon}

\newcommand{\go}{\omega}

\newcommand{\q}{\,,\qquad}

\newcommand{\nn}{\nonumber}

\newcommand{\half}{\frac{1}{2}}

\newcommand{\p}{\partial}

\newcommand{\f}{\frac}

\newcommand{\ups}{\upsilon}
\newcommand{\bu}{\bar{\kappa}}

\def\ck{{\cal K}}

\newcommand{\PPP}{ {J} }

\newcommand{\xx}{{\bf x}}

\newcommand{\dgb}{{\dot \gb}}
\newcommand{\dga}{{\dot \ga}}
\newcommand{\dr}{{\rm d}}
\newcommand{\zz}{{\bf z}}
\newcommand{\Bb}{{\mathcal B}}
\begin{document}

\begin{flushright}
{\small FIAN/TD/04-16}
\end{flushright}
\vspace{1.7 cm}

\begin{center}
{\large\bf Symmetries and Invariants in Higher-Spin Theory}

\vspace{1 cm}

{\bf  M.A.~Vasiliev}\\
\vspace{0.5 cm}
{\it
 I.E. Tamm Department of Theoretical Physics, Lebedev Physical Institute,\\
Leninsky prospect 53, 119991, Moscow, Russia}

\end{center}

\vspace{0.4 cm}

\begin{abstract}
\noindent
General aspects of  higher-spin gauge theory and unfolded formulation
are briefly recalled with some emphasize on the recent results on the  
breaking of $sp(8)$ symmetry by current interactions and  construction 
of invariant functionals relevant to the higher-spin holography.

\end{abstract}

\newpage
\tableofcontents

\newpage

\section{Introduction}
Higher-spin (HS) gauge theory is based on higher symmetries associated with
HS massless fields. HS gauge symmetries are expected to become manifest
at ultra high energies possibly beyond the Planck energy.
Since such energies are  unreachable  by modern accelerator devices
 the conjecture that a fundamental theory  exhibits
the HS symmetries at ultra high energies  provides a unique
chance to explore properties of this regime. HS symmetries severely
restrict the structure of HS theory.

The study of HS fields has long history  starting from  seminal
papers of Dirac \cite{Dirac:1936tg}, Fierz and Pauli \cite{Fierz:1939ix},
and others including the Tamm group
\cite{42,GT,fr}. The role of HS gauge
symmetries for massless fields in four dimensions
was originally appreciated at the linearized level for spin 3/2
by Rarita and Schwinger \cite{Rarita:1941mf}
and for any spin  by Fronsdal  \cite{Fronsdal:1978rb}.

Extension to the interacting level was not simple encountering difficulties of
combining   nonAbelian symmetries of different types.
First positive results were obtained in eighties of the last century in the papers by
A.~Bengtsson, I.~Bengtsson, Brink \cite{Bengtsson:1983pd,Bengtsson:1983pg}
and  Berends, Burgers, van Dam
\cite{Berends:1984wp,Berends:1984rq}
who found that the action consistent with HS gauge symmetries
in the cubic order contains higher derivatives in interactions
$$S=S^2+S^3+\dots \q S^3=\sum_{p,q,r}(D^p \varphi)(D^q
\varphi)(D^r \varphi)\rho^{p+q+r+\half d-3}\,,
$$
where the order of higher derivatives increases with spins of the fields
$\varphi$  in the vertex. Since full HS theory necessarily
involves infinite towers of HS fields, such a theory is somewhat nonlocal
(note however that no higher derivatives appear at the quadratic level
within the expansion around $AdS$ background). Of course some kind of nonlocality
beyond Planck scale should be expected of the theory anticipated to capture the
quantum gravity regime.

Appearance of higher derivatives in interactions
 requires a dimensionful coupling constant $\rho$ in the action, whose
origin in a theory  involving only  massless fields was obscure.
Resolution of this puzzle was proposed in \cite{Fradkin:1987ks,Fradkin:1986qy}
via identification of the parameter $\rho$   with the radius of the background
(anti-)de Sitter space. In this setup, the higher derivative  vertices do not allow
a meaningful flat limit. This is consistent with numerous no-go statements
ruling out consistent interactions of massless HS fields in Minkowski space
\cite{Coleman:1967ad,Aragone:1979hx}.

In eighties, the fact that consistent HS interactions require
non-zero cosmological constant looked like a peculiarity. It
acquired a much deeper interpretation after discovery of the $AdS/CFT$
correspondence \cite{Maldacena:1997re,Gubser:1998bc,Witten:1998qj}.
The fact that
HS theories are most naturally formulated in the $AdS$ background
was conjectured  to play a role in the context of the HS
holography in \cite{Sundborg:2000wp,WJ,Sezgin:2002rt}. The precise
conjecture on the $AdS_4/CFT_3$
correspondence was put forward by Klebanov and Polyakov
\cite{Klebanov:2002ja} while its first explicit check
 was performed by Giombi and Yin in \cite{Giombi:2009wh}. This research triggered
a great increase of interest in HS theories and HS holographic duality
(see e.g.
\cite{Maldacena:2012sf}-\cite{Hegde:2015dqh} and references therein).
$CFT_2$ duals of $3d$ HS symmetries were studied in
\cite{Henneaux:2010xg,Campoleoni:2010zq}. The conjecture
on $AdS_3/CFT_2$ HS holography proposed by Gaberdiel and Gopakumar
\cite{Gaberdiel:2010pz} also
formed an active research direction in the recent years.

One of the  driving forces of the study of interacting HS theories
 is the hope that HS holography
may help to uncover the origin of $AdS/CFT$. However, the subtlety is  that
 despite significant progress in the construction of actions during
last thirty years \cite{Bengtsson:1983pd}-\cite{Fradkin:1986qy},
\cite{Koh:1986vg}-\cite{Joung:2012hz}
the full nonlinear generalization of the Fronsdal action is still
unavailable. This complicates the standard construction of
the generating functional. One of the aims of
this talk is to discuss a recently proposed construction for
invariant on-shell functionals \cite{Vasiliev:2015mka}  conjectured to give
rise to both the generating functional for boundary correlators and black hole
charges.

\section{Unfolded Dynamics}
\label{Unfolded Dynamics}
\subsection{General setup}
Modern formulation of nonlinear HS theory \cite{more} is based on the so-called
unfolded approach \cite{Ann} which reformulates dynamics in question in terms
of differential forms.
{The unfolded form of dynamical equations provides a covariant generalization
of the first-order form of  differential equations}
$$
\dot{q}^i(t) =\varphi^i (q(t))\,,
$$
which is convenient in many respects. In particular, since
initial data can be given in terms of the values of variables
$q^i (t_0 )$ at any given point $t_0$,  the number of degrees of freedom
in the first-order formulation equals to the number of dynamical variables.

{Unfolded formulation is a multidimensional generalization achieved via
 replacement of the time derivative by the de Rham derivative}
$$\f{\p}{\p t} \to \dr= dx^\nu \p_\nu$$
and the dynamical variables $q^i$ by a set of differential forms
$$
 q^i(t)\rightarrow W^\Omega (dx,x)=dx^{\nu_1}\ldots dx^{\nu_p}
W^\Omega_{\nu_1 \ldots \nu_p} (x) $$
to reformulate a system of partial differential equations in the
first-order covariant form
\be
\label{unf}
\dr W^\Omega (dx,x)=G^\Omega (W(dx,x))\,.\qquad
\ee
Here $dx^\nu$ are anticommuting differentials
($dx^\nu  dx^\mu = - dx^\mu dx^\nu$; the wedge product symbol is implicit) and
$G^\Omega (W)$ {are some functions of the ``supercoordinates"}
$W^\Omega$
$$
G^\Omega (W) = \sum_n f^\Omega{}_{\Lambda_1\ldots
\Lambda_n}W^{\Lambda_1}  \ldots W^{\Lambda_n}\,.
$$
Since $\dr^2 =0$ at $d>1$ {the functions $G^\Lambda (W)$  have to obey
the  compatibility conditions}
\be
\label{cc}
 G^\Lambda (W)\f{\p
G^\Omega (W)} {\p W^\Lambda}  \equiv 0\,.
\ee
(Not that all products of the differential forms $W(dx,x)$ are the wedge
products due to anticommutativity of  $dx^\nu$.)
Let us stress that these are conditions on the functions $G^\Lambda (W)$ rather than on
$W$.

As a consequence of (\ref{cc}) system (\ref{unf}) is  invariant under
the gauge transformation
\be\label{gtr} \delta W^\Omega = d \varepsilon^\Omega +\varepsilon^\Lambda
\frac{\p G^\Omega (W) }{\p W^\Lambda}\,,
\ee
 {where the gauge parameter} $\varepsilon^\Omega (x) $ {is a}
$(p_\Omega -1)$-form for a $p_\Omega$-{form} $W^\Omega$. {Strictly
speaking, this is true for the class of {\it universal} unfolded systems in which
the compatibility conditions (\ref{cc}) hold independently of the dimension  $d$ of
space-time, \ie (\ref{cc}) should be
true disregarding the fact that any $(d+1)$-form is zero. Let us stress that
all unfolded systems appearing in HS theories are universal.}

The unfolded formulation can be applied to description of
invariant functionals.
As shown in \cite{Vasiliev:2005zu}, the variety of gauge invariant functionals
associated with the unfolded equations (\ref{unf}) is described by the cohomology
of the operator
\be
\label{Q}
Q= G^\Omega(W) \f{\p}{\p W^\Omega}\,,
\ee
which obeys
$$
Q^2=0
$$
as a consequence of (\ref{cc}). By virtue of (\ref{unf}), $Q$-closed $p$-form functions
$L_p(W)$ are $\dr$-closed, giving rise to the gauge invariant functionals
$$
S=\int_{\Sigma^p} L_p\,.
$$
In the off-shell case they can be used to construct invariant actions
while in the on-shell case they describe conserved charges. (For more detail and
examples see \cite{Vasiliev:2005zu}.)

\subsection{Properties}

The unfolded formulation of partial differential equations has a number of
remarkable properties.

\begin{itemize}
\item

First of all, it has general applicability:
 every system of partial differential equations can be
reformulated in the unfolded form.

\item

{Due to using the exterior algebra formalism, the system
is invariant under diffeomorphisms, being coordinate independent.}

\item

 Interactions can be understood as  nonlinear deformations of $G^\Omega(W)$.

\item

{Unfolded formulation gives  clear group-theoretical interpretation   of
fields and equations in terms of modules and Chevalley-Eilenberg cohomology
of a  symmetry algebra} $h$. In particular,  background fields are described by
a flat connection of  $h$. {Dynamical fields are described in terms of} $h${-modules
while equations of motion have a form of covariant constancy conditions.}

\item
Local degrees of freedom are carried by 0-forms
$C^i(x_0)$ {at any} {$x=x_0$} (as $q(t_0)$), valued in the
infinite-dimensional module dual to the
space of single-particle states: $C^i(x_0)$ are moduli of solutions
replacing initial data in the conventional Cauchy problem.
{It is worth to mention that this property of  unfolded dynamics gives a
tool to control unitarity in presence of higher derivatives via the requirement on
the space of zero-forms like $C^i(x_0)$ to admit a positive-definite norm
preserved by the unfolded equations in question.}

\item
The most striking feature of the unfolded formulation is
that it makes it possible to describe one and the same dynamical system in
space-times of different  dimensions.
{Unfolded dynamics exhibits independence of the ``world-volume" space-time with
coordinates $x$.}
{Instead, geometry is encoded by} the functions $G^\Omega (W)$ in the ``target space"
of fields $W^\Omega$.
{Indeed, the universal unfolded equations make sense in any space-time independently of
a particular realization of the de Rham derivative $\dr$. For instance one can
extend space time by adding additional coordinates $z$}
$$
 \dr W^\Omega (x)=G^\Omega (W(x))\,,\quad x\rightarrow X=(x,z)\,,\quad
\dr_x\rightarrow \dr_X = \dr_x +\dr_z\,,\quad \dr_z =
dz^u\f{\p}{\p z^u}\,.
$$
The unfolded equations reconstruct the
$X${-dependence in terms of values of the fields} $W^\Omega(X_0)=W^\Omega(x_0,z_0)$
{at any} $X_0$. {Clearly, to take} $W^\Omega(x_0,z_0)$ in space $M_X$
{with coordinates} $X_0$ {is the same as to take} $W^\Omega(x_0)$ {in the space}
$M_x\subset M_X$  with coordinates  $x$.

Such uplifting becomes most interesting
provided that there is a nontrivial vacuum connection along the additional
coordinates $z$. This is in particular the case for $AdS/CFT$ correspondence where
the conformal flat connection at the boundary is extended to the flat $AdS$
connection in the bulk with $z$ being a radial Poincar\'e coordinate.

\end{itemize}

\section{Field-current-field correspondence}
The $AdS_4/CFT_3$ HS holography \cite{Klebanov:2002ja} relates the HS
gauge theory in $AdS_4$ to the quantum theory of conformal currents in
three dimensions. To see how it works from the unfolded dynamics perspective,
let us first discuss the unfolded equations for free massless fields
and currents on the $3d$ boundary.

The unfolded equations of $3d$ conformal  massless fields is
formulated in terms of 0-forms  $C(y|x)$ \cite{Shaynkman:2001ip,Vasiliev:2001zy}
 which depend on the coordinates $\xx^{\ga\gb}=\xx^{\gb\ga}$ of $3d$ space-time
and auxiliary commuting spinorial variables $y^\ga$ ($\ga,\gb =1,2$ are $3d$
spinorial indices). Unfolded equations for conformal  massless fields
are
\be
\label{3dun}
(\f{\p}{\p \xx^{\ga\gb}} \pm i \f{\p^2}{\p y^\ga \p y^\gb} )
C_j^\pm(y|\xx)=0\q \ga,\gb=1,2\,,\quad
\ee
where $ j=1,\ldots {\mathcal N}$ is a color index.

The meaning of these equations is twofold. Firstly, they express all terms of degree
 two or higher in the $y$ variables via $x$-derivatives of the fields $C^\pm(\xx):=
 C^\pm (0|\xx)$
 and $C^\pm_\ga(\xx):=\f{\p}{\p y^\ga}C^\pm (y|\xx)\Big |_{y=0}$. The latter  are the usual scalar and
 spinor fields which obey,
 respectively,  Klein-Gordon and Dirac equations by virtue of
 (\ref{3dun}).  More precisely, the $\pm$ components should be identified with
 positive- and negative-frequency parts of the solutions of free field equations.
 Note that the fields
$C^\pm(\xx)$ and $C^\pm_\ga(\xx)$ are primaries of the conformal modules underlying
 equations   (\ref{3dun}).

The unfolded equations for $3d$ conformal conserved currents have the rank-two form
\cite{Gelfond:2003vh}
\be \label{xyy}\left\{
\,\f{\p}{\p \xx{}^{\ga\gb}}  -  \f{\p^2}{\p y^{(\ga}\p u^{\gb)}}
 \right\} J(u,\,y| \xx) =0\,.
 \ee
$J(u,\,y| \xx)$ {contains all $3d$ HS currents along with their
derivatives.

Elementary $3d$ conformal currents}, which are conformal {primaries},
{contain conserved currents of all spins}
\be\nn
J(u,0|\xx)= \sum_{2s=0}^\infty u^{\ga_1}\ldots u^{\ga_{2s}}
J_{\ga_1\ldots \ga_{2s}}(\xx)\,,\quad \tilde
J(0,y|\xx)= \sum_{2s=0}^\infty y^{\ga_1}\ldots y^{\ga_{2s}}
\tilde J_{\ga_1\ldots \ga_{2s}}(\xx)\,
\ee
along with the additional scalar current
$$
J^{asym}(u,y|\xx) = u_\ga y^\ga J^{asym} (\xx)\,.
$$
Their conformal dimensions are
$$
\Delta J_{\ga_1\ldots \ga_{2s}}(\xx) = \Delta \tilde
J_{\ga_1\ldots \ga_{2s}}(\xx)= s+1
\qquad
\Delta J^{asym}(\xx)=2\,.
$$

{The unfolded equations express all other components of $J(u,y|\xx)$ in terms of
derivatives of the primaries, also imposing the differential equations on the
latter, which are just the conservation conditions
\be\nn
\f{\p}{\p \xx^{\ga\gb}}\f{\p^2}{\p u_\ga \p u_\gb} J(u,0|\xx) =0\q
\f{\p}{\p \xx^{\ga\gb}}\f{\p^2}{\p y_\ga \p y_\gb} \tilde J(0,y|\xx) =0\,
\ee
for all currents  except for the
scalar ones that obey no differential equations}.

The rank-two equation is obeyed by
\be\nn
J(u,\,y\,|\xx) =\sum_{i=1}^{\mathcal N}
C^-_{i}( {u}+y|\xx)\, C^+_{i}( y-u|\xx)\,.
\ee
This simple formula gives the explicit realization of the HS conformal
conserved currents in terms of bilinear combinations of  derivatives of
free massless fields in three dimensions.

{Generally, the rank-two fields and hence conserved currents can be interpreted as
bi-local fields in the twistor space.} In this respect they are somewhat
analogous to space-time bi-local fields also used for the description
of currents (see e.g \cite{Jevicki:2012fh} and references
therein).

To relate $3d$ currents to $4d$ massless fields we extend the
$3d$ current equation to the $4d$ {massless equations. This is easy to
achieve in the
 unfolded dynamics via the extension of the $3d$ coordinates} $\xx^{\ga\gb}$  to
the $4d$ coordinates $x^{\ga\dgb}$, extending  $3d$ equations (\ref{xyy}) to
\be
\label{Xyy}
\left ( \f{\p}{\p x^{\ga \dga}} + \f{\p^2}{\p y^\ga \p \bar y^\dgb} \right )
C(y,\bar y|x) =0\,.
\ee
{These are just the free unfolded equations \cite{Ann} for} $4d$  {massless
fields of all spins in Minkowski space, \ie at} $\Lambda=0$.

The analysis in $AdS_4$, which is also simple, is performed analogously.
In this case,
$
\xx^{\ga\gb} = \half ( x^{\ga \gb} + x^{\gb \ga})
$
{are boundary coordinates}, while
 $z^{-1} = x^{\ga\gb}\epsilon_{\ga\gb}$ is the {radial coordinate}.
 (For more detail see \cite{Vasiliev:2012vf}.)
 {At the non-linear level, the full HS theory in} $AdS_4$ {turns out to be
  equivalent to the theory  of } $3d$ {currents of all spins interacting
  through conformal HS gauge fields \cite{Vasiliev:2012vf}.}

{A rank-two field (current) in} $AdS_3$ is equivalent to
a rank-one field in a larger  space with ten coordinates $X^{AB}=X^{BA}$
$$
(\f{\p}{\p X^{AB}} + \f{\p^2}{\p y^A \p y^B} )
J^3(y|x)=0\q A,B=1,\ldots \,, 4\q X^{AB}=X^{BA}\,,
$$
$$X^{AB}= (x^{\ga\dga}, x^{\ga\gb}, \bar x^{\dga\dgb})\q
x^{\ga\dga} = ( {\bf x^{\ga\dga}} , \gvep^{\ga\dga} {\bf z} )\,.
$$
{Reduction to Minkowski coordinates} $x^{\ga\dga}$ {gives }
$4d$ {massless equations for all spins with}
$
J^3 \to C^4\,.
$
Mathematically this is the manifestation of the Flato-Fronsdal  theorem \cite{FF}
stating that the tensor product of unitary modules associated with $3d$ massless
fields gives the unitary module associated with all $4d$ massless fields:
\be
(3d, m=0)\otimes (3d, m=0)=\sum_{s=0}^\infty (4d,
m=0)\,.
\ee

{The full system of $4d$ massless fields of
all spins exhibits} $sp(8)$ {symmetry}
\cite{Fr1,BL,BLS,Vasiliev:2001zy}. {A rank-two field in}  four dimensions
{describes} $4d$ {conserved currents equivalent to a rank-one
field in six dimensions \cite{BLS,Mar}
$$
C^4 C^4 \sim J^4 \sim C^6\,.
$$
Dualities of this type can be called field-current-field correspondence.

\section{From free massless equations to current interactions and holography}

\subsection{Central on-shell theorem}
The infinite set of $4d$ massless fields of all spins $s=0,1,2\ldots $ is
conveniently described by a
{1-form} $  \omega (y,\bar{y}\mid x)\,,$ and 0-form $ C(y,\bar{y}\mid x)  $
$$
 A
(y,\bar{y}\mid x) =i\sum_{n,m=0}^{\infty} \frac{1}{n!m!}
{y}_{\alpha_1}\ldots {y}_{\alpha_n}{\bar{y}}_{{\pb}_1}\ldots
{\bar{y}}_{{\pb}_m } A{}^{\alpha_1\ldots\alpha_n}{}_,{}^{{\pb}_1
\ldots{\pb}_m}(x)\,.
$$
{The central fact of the analysis of free massless fields in four dimensions
known as Central on-shell theorem is that unfolded system for free massless
fields has the form \cite{Ann}}
\bee
\label{CON1}
    && R_1(y,\overline{y}\mid x) =
    \overline{H}^{\dga\pb}
\f{\p^2}{\p \overline{y}^{\dga} \p \overline{y}^{\dgb}}\
{\bar C}(0,\overline{y}\mid x) + H^{\ga\gb} \f{\p^2}{\p
{y}^{\ga} \p {y}^{\gb}}\
{C}(y,0\mid x)\,, \\\label{CON2}
\,&& \tilde{D}_0C (y,\overline{y}\mid x) =0\,, \eee
where
\be
\label{RRR}
R_1 (y,\bar{y}| x) =D^{ad}\omega(y,{\bar{y}}|x) =
D^L \omega (y,\bar{y}| x) -
\lambda e^{\ga\pb}\Big (y_\ga \frac{\partial}{\partial \bar{y}^\pb}
+ \frac{\partial}{\partial {y}^\ga}\bar{y}_\pb\Big )
\omega (y,\bar{y} | x) \,,
\ee
\be
\label{tw}
\tilde D C(y,{\bar{y}}|x) =
D^L C (y,{\bar{y}}|x) +\f{i}{2}\lambda e^{\ga\pb}
\Big (y_\ga \bar{y}_\pb -\frac{\partial^2}{\partial y^\ga
\partial \bar{y}^\pb}\Big ) C (y,{\bar{y}}|x)\,,
\ee
\be
\label{dlor}
D^L A (y,{\bar{y}}|x) =
d A (y,{\bar{y}}|x) -
\Big (\go^{\ga\gb}y_\ga \frac{\partial}{\partial {y}^\gb} +
\overline{\go}^{\pa\pb}\bar{y}_\pa \frac{\partial}{\partial \bar{y}^\pb} \Big )
A (y,{\bar{y}}|x)\,.
\ee
Here the background $AdS_4$ Lorentz connection
$\omega_{\alpha \gb} $, $\overline{\omega}_{\dga\dgb}$ and
vierbein  $e_{\ga\dgb}$ obey the $AdS_4$  equations
\be
\label{adsfl}
R_{\ga\gb}=0\,,\quad \overline{R}_{\pa\pb}=0\,,
\quad R_{\ga\pa}=0\,,
\ee
where $\lambda^{-1}$ is the  $AdS_4$ radius and
\be
\label{nR}
R_{\alpha \gb}=d\omega_{\alpha \gb} +\omega_{\alpha}{}^\gamma
\wedge \omega_{\gb \gamma} +\lambda^2\, e_{\alpha}{}^{\dot{\delta}}
\wedge
e_{\gb \dot{\delta}}\,,
\ee
\be
\nn
\overline{R}_{{\pa} {\pb}}
=d\overline{\omega}_{{\pa}
{\pb}} +\overline{\omega}_{{\pa}}{}^{\dot{\gamma}}
\wedge \overline{\omega}_{{\pb} \dot{\gga}} +\lambda^2\,
e^\gamma{}_{{\pa}} \wedge e_{\gamma {\pb}}\,,
\ee
\begin{equation}
\label{nr}
R_{\alpha {\pb}} =de_{\alpha{\pb}} +\omega_\alpha{}^\gamma \wedge
e_{\gamma{\pb}} +\overline{\omega}_{{\pb}}{}^{\dot{\delta}}
\wedge e_{\alpha\dot{\delta}}\,.
\end{equation}
(Two-component indices are raised and lowered
by $\varepsilon_{\ga\gb}$ or $\varepsilon_{\pa\pb}$.)
$H^{\ga\gb}=H^{\gb\ga}$ and $\overline{H}^{\pa\pb} =
\overline{H}^{\pb\pa}$ are the basis 2-forms
\be
\label{H}
H^{\ga\gb} := e^{\ga}{}_\pa  e^\gb{}^\pa\,,\qquad
\overline {H}^{\pa\pb} := e_{\ga}{}^\pa e^{\ga\pb}\,.
\ee

The {0-forms} $C(Y|x)$ {form a Weyl module equivalent to the  boundary
current module.} 1-form HS connections $  \omega (y,\bar{y}\mid x)$
contain HS gauge fields. For spins $s\geq 1$, equation (\ref{CON1})
expresses the Weyl {0-forms} $C(Y|x)$ via
gauge invariant combinations of derivatives of the HS gauge connections.
From this perspective the Weyl {0-forms} $C(Y|x)$ generalize the spin-two
Weyl tensor along with all its derivatives to any spin.

\subsection{Current deformation}
{Schematically,} {for the flat connection} $D=\dr+w$ the current deformation
of the free equations (\ref{CON1}), (\ref{CON2}) has the form
  \bee\nn\left\{\beee{l}
  D\go^4+L(C^4,w)=0\\
  \tilde DC^4=0\\
  D_2 \PPP^4=0\eeee\right.\quad {\Rightarrow}\quad
  \left\{\beee{l}
  D\go^4+L(C^4,w)+G(w ,\PPP^4)=0\\  \tilde D C^4+F(w,\PPP^4)=0\\
   D_2  \PPP^4=0\eeee\right.
\overline{}  \eee

   {The sector of 0-forms} of this system was analyzed in detail in
\cite{Gelfond:2010pm,Gelfond:2015poa}.
Here $J^4$ {can be interpreted either as a} $4d$ {current or as
a} $6d$ {massless field}.
As a result, $4d$ {current interactions can be interpreted as a  mixed linear
system of} $d4$ {and} $d6$
{fields \cite{Gelfond:2010pm}.}
{Algebraically this is the semidirect sum  of a rank-one and rank-two systems.}

{An interesting question  is what symmetry is preserved by the deformed system?}
{When unmixed, both rank-one and rank-two system are} $sp(8)${-invariant.
The question whether} $sp(8)$ {is preserved by the deformation
is equivalent to that whether formal consistency of the deformation
takes place with} any connection $w\in sp(8)$. The analysis of this question
\cite{Gelfond:2015poa} shows that
{current interactions break} $sp(8)$ {down to the conformal algebra} $su(2,2)$.

\subsection{Kinematics of $AdS_4 /CFT_3$ HS holography}

{To make boundary conformal invariance manifest it is convenient to use
the following basis}
\be\nn
y^+_\ga = \half (y_\ga - i \bar y_\ga)\q
y^-_\ga =\half (\bar y_\ga -i y_\ga)\q
[ y^-_\ga\,, y^{+\gb }]_\star{}  = \delta_{\ga}^{\gb}\,.
\ee
 $AdS_4$ {can be foliated as}
$
x^{{n}}=({\bf x}^{{a}},\zz)\,,
$
where $\xx^{{a}}$ {are coordinates of leaves} (${a}=0,1,2$,) and the
{Poincar\'{e} coordinate}  $\zz$
{is the foliation parameter.} $AdS_4$ {infinity is at} $\zz=0$. In these
coordinates the background connection at $\lambda=1$  is
\be\nn
\label{w0}
W =\f{i}{ \zz}d\xx^{\ga\gb} y^-_\ga y^-_\gb - \f{d\zz}{2\zz} y^-_\ga\,
y^{+\ga}\,,
\ee
\be\label{poframe}\nn
e^{\ga\dga} = \f{1}{2\zz} dx^{\ga\dga}\q \go^{\ga\gb}=-
\f{i}{4\zz} d\xx^{\ga\gb}\q \bar \go^{\dga\dgb} =
\f{i}{4\zz} d\xx^{\dga\dgb}\,.
\ee

{Using insensitiveness of  unfolded equations to the extension to a larger
space, the vacuum connection can be analitically  extended to the complex plane of} $\zz$ {
with all components containing} $d\bar \zz$ {being zero}.
{In these terms the generating functional for the boundary correlators
takes the form}
{
\be\nn
\label{S}
S=\f{1}{2\pi i}\oint_{\zz=0} \Ll(\go(C),C)
\ee
}
{if $\Ll(\go(C),C)$ is an on-shell closed} $(d+1)${-form}  {for
 a} $d${-dimensional boundary}
\be\nn
\dr \Ll(\go(C),C)=0\q \Ll \neq \dr M\,.
\ee
The resulting {functional} {is the residue at}
$\zz=0$ {giving the  boundary functional of the
structure analogous to} $ \phi_{n_1 \ldots n_s} J^{n_1 \ldots n_s}$
\be\nn \label{freel} S_{M^3}(\go)=\int_{M^3} \Ll \q \Ll=\half
\go_\xx^{\ga_1\ldots \ga_{2(s-1)}} e_\xx^{\ga_{2s-1}}{}_\gb
e_{\xx}^{\ga_{2s}}{}^\gb ( a C_{\ga_1\ldots \ga_{2s}}(\go)+\bar a
\bar C_{\ga_1\ldots \ga_{2s}}(\go))\,. \ee
Here $C_{\ga_1\ldots\ga_{2s}}(\go)$, which {have conformal properties of currents
$J$,
are expressed via the HS connections $\go$ by Eq.~(\ref{CON1}).} On the other hand
 $\omega^{\ga_1\ldots \ga_{2(s-1)}}$ {have conformal dimensions of the
shadow sources $\phi$ to the currents.
Being related to $C $ {via unfolded equations} it
does not describe new degrees of freedom however.

The $C$-dependent terms can be represented in the form
$$
a C_{\ga_1\ldots \ga_{2s}}(\go)+\bar a \bar C_{\ga_1\ldots \ga_{2s}}(\go)
=a_- {\mathcal T}_{-\ga_1\ldots \ga_{2s}}(\go)+
a_+ {\mathcal T}_{+\ga_1\ldots \ga_{2s}}(\go)\,,
$$
where
$\mathcal T_-$ {describes local boundary terms} while
$\mathcal T_+$ {describes nontrivial correlators via the
variation of} $S_{M_3}$ {over the HS gauge fields}
$\go_\xx^{\ga_1\ldots \ga_{2(s-1)}}$ \be\nn \langle
J(\xx_1)J(\xx_2)\ldots \rangle = \f{\delta^n \exp{[-  S_{M^3}}(\go,
C(\go))]}{\delta \go(\xx_1) \delta\go(\xx_2)\ldots} \Big |_{\go=0}\,.
\ee

The main problem is to find  an appropriate nonlinear invariant functional $\Ll$.

\section{Nonlinear HS equations in $AdS_4$}
\label{Nonlinear Higher-Spin Equations}

To explain the construction of invariant functionals we first recall the
form of nonlinear massless field equations of \cite{more}.
The key element  is the doubling of auxiliary
Majorana spinor variables $Y_A$ in the HS 1-forms and 0-forms
\be
\go(Y;\ck|x)\longrightarrow W(Z;Y;\ck|x)\,,\qquad
C(Y;\ck|x)\longrightarrow B(Z;Y;\ck|x)
\ee
supplemented with equations which determine  dependence on the
additional variables $Z_A$ in terms of ``initial data"
\be
\label{inda}
\go(Y;\ck|x)=W(0;Y;\ck|x)\,,\qquad C(Y;\ck|x)= B(0;Y;\ck|x).
\ee
An additional spinor field $S_A (Z;Y;\ck|x)$, that carries only pure gauge
degrees of freedom, plays a role of connection in
 $Z^A$ directions. It is convenient to
introduce anticommuting $Z-$differentials $dZ^A dZ^B=-dZ^B
dZ^A$ to interpret $S_A (Z;Y;\ck|x)$ as a $Z$--1-form,
\be
S=dZ^A S_A (Z;Y;\ck|x) \,.
\ee
The variables $\ck=(k,\bar{k})$ are Klein operators
that satisfy
\be
\label{kk}
k w^\ga = -w^\ga k\,,\quad
k \bar w^\pa = \bar w^\pa k\,,\quad
\bar k w^\ga = w^\ga \bar k\,,\quad
\bar k \bar w^\pa = -\bar w^\pa \bar k\,,\quad k^2=\bar k^2 = 1\,,\quad
k\bar k = \bar k k\,
\ee
 with
 $w^\ga= (y^\ga, z^\ga, dz^\ga )$, $\bar w^\pa =
(\bar y^\pa, \bar z^\pa, d\bar z^\pa )$.

The nonlinear HS equations are \cite{more}
\be
\label{dW}
dW+W*W=0\,,\qquad
\ee
\be
\label{dB}
dB+W*B-B*W=0\,,\qquad
\ee
\be
\label{dS}
dS+W*S-S*W=0\,,
\ee
\be
\label{SB}
S*B=B*S\,,
\ee
\be
\label{SS}
S*S= -i (dZ^A dZ_A + dz^\ga dz_\ga  F_*(B) k\ups +
d\bar z^\dga d\bar
z_\dga \bar F_*(B) \bar k \bu)
\,,
\ee
where
$F_*(B) $ is some star-product function of the field $B$.

Setting
$
\W= \dr +W +S
$
brings equations (\ref{dW})-(\ref{SS}) to the concise form
\be
\W *{} \W = -i[ dZ_A dZ^A + \eta \delta^2(dz)  \Bb * k*\kappa +
\bar \eta \delta^2( d{\bar z}) \Bb * \bar k * \bar\kappa]\,,
\ee
\be
\W*{} B = B*{}\W\,.
\ee

The simplest choice  of linear functions
\be
\label{etaB}
F_*(B)=\eta B \q \bar F_* (B) = \bar\eta B\,,
\ee
where $\eta$ is some phase factor (its absolute value can be absorbed into
redefinition of $B$) leads to a class of pairwise nonequivalent nonlinear HS
theories. The particular cases
of $\eta=1$ and $\eta =\exp{\f{i\pi}{2}}$ are especially interesting, corresponding
to so called $A$ and $B$ HS models. These two cases are distinguished
by the property that they  respect parity \cite{Sezgin:2003pt}.

The associative star product $*$ acts on functions of two
spinor variables
\be
\label{star2}
(f*g)(Z;Y)=\frac{1}{(2\pi)^{4}}
\int d^{4} U\,d^{4} V \exp{[iU^A V^B C_{AB}]}\, f(Z+U;Y+U)
g(Z-V;Y+V) \,,
\ee
where
$C_{AB}=(\varepsilon_{\ga\gb},  \varepsilon_{\dga\dgb})$
is the $4d$ charge conjugation matrix and
$ U^A $, $ V^B $ are real integration variables. It is
normalized so that 1 is a unit element of the star-product
algebra, \ie $f*1 = 1*f =f\,.$ Star product
(\ref{star2}) provides a particular
realization of the Weyl algebra
\be
[Y_A,Y_B]_*=-[Z_A,Z_B ]_*=2iC_{AB}\,,\qquad
[Y_A,Z_B]_*=0\,
\ee
($[a,b]_*=a*b-b*a$).

The left and right inner Klein operators
\be
\label{kk4}
\kappa =\exp i z_\ga y^\ga\,,\qquad
\bu =\exp i \bar{z}_\dga \bar{y}^\dga\,,
\ee
 which enter Eq.~(\ref{SS}), change a sign of
 undotted and dotted spinors, respectively
\be
\label{uf}
\!(\kappa *f)(z,\!\bar{z};y,\!\bar{y})\!=\!\exp{i z_\ga y^\ga }\,\!
f(y,\!\bar{z};z,\!\bar{y}) ,\quad\! (\bu
*f)(z,\!\bar{z};y,\!\bar{y})\!=\!\exp{i \bar{z}_\dga \bar{y}^\dga
}\,\! f(z,\!\bar{y};y,\!\bar{z}) ,
\ee
\be
\label{[uf]}
\kappa *f(z,\bar{z};y,\bar{y})=f(-z,\bar{z};-y,\bar{y})*\kappa\,,\quad
\bu *f(z,\bar{z};y,\bar{y})=f(z,-\bar{z};y,-\bar{y})*\bu\,,
\ee
\be
\kappa *\kappa =\bu *\bu =1\q \kappa *\bu = \bu*\kappa\,.
\ee

To analyze Eqs.~(\ref{dW})-(\ref{SS}) perturbatively,
one has to linearize them around some vacuum solution.
The simplest choice is
\be
\label{vac}
W_0(Z;Y|x)= W_0(Y|x)\q S_0(Z;Y|x) = dZ^A Z_A\q B_0=0\,,
\ee
where $W_0(Y|x)$ is some solution of the flatness condition
\be
dW_0(Y|x) + W_0(Y|x)*W_0(Y|x)=0\,.
\ee
 $W_0(Y|x)$ bilinear in $Y^A$ describes $AdS_4$.

 Propagating massless fields are described by the fields
$\W(Z;Y;\ck|x)$ even in $\ck$ and fields $B(Z;Y;\ck|x)$ odd
in $\ck$
\be
\W(Z;Y;-\ck|x)=\W(Z;Y;\ck|x)\q B(Z;Y;-\ck|x)=-B(Z;Y;\ck|x)\,.
\ee
In this sector, linearization of  system
(\ref{dW})-(\ref{SS}) around  vacuum (\ref{vac}) just
reproduces  free  field equations (\ref{CON1}), (\ref{CON2}).

The fields of opposite parity in the Klein operators
\be
\W(Z;Y;-\ck|x)=-\W(Z;Y;\ck|x)\q B(Z;Y;-\ck|x)=B(Z;Y;\ck|x)\,
\ee
are topological in the sense that irreducible fields describe at most a
finite number of degrees of freedom.
(For more detail see \cite{more,Vasiliev:1999ba,Didenko:2014dwa}).
As such they can be treated as describing infinite sets of the coupling constants
in HS theory.

\section{Invariants of the $AdS_4$ HS theory}
\label{Invariants}
To explain the idea of our construction let us first consider
an example of  {a contractible unfolded system
of the form}
\be
\label{Conteq}
\dr w = \Ll\q \dr \Ll=0\,.
\ee
It is obviously consistent and hence is invariant under
 gauge transformations (\ref{gtr})
\be
\label{trg}
\delta w(x) = \epsilon (x)\q \delta \Ll (x) = \dr \epsilon (x)\,.
\ee
As such it is dynamically empty since the gauge transformation
allows one to  {gauge fix} $w=0$. By virtue of (\ref{Conteq}) it follows
then that  $ \Ll=0$.

{A more interesting system is}
\be
\label{L}
\dr w +L(W)= \Ll\q \dr \Ll=0\,,
\ee
{where} $L(W)$ {is some closed function of other fields} $W$
that obey some unfolded equations (\ref{unf}).
{In the {\it canonical gauge}} $w=0$ it takes the form
$$
\Ll=L(W)\q \dr L(W)=0\,.
$$
{The singlet field} $L$ {becomes a Lagrangian
giving rise to an invariant action}
\be
\label{S}
S=\int_\Sigma L(W)\,.
\ee
So defined functional is independent of local variations of the
integration cycle and gauge invariant. Indeed, being formally consistent,
the system is invariant under gauge transformations (\ref{gtr}) with respect to
the gauge parameter $\epsilon$ associated with $w$ and the gauge
parameters $\varepsilon^\Omega$ associated with $W^\Omega$. In the gauge
$w=0$, the parameter $\epsilon$ is expressed by the condition $\delta w=0$
via the gauge parameters $\varepsilon^\Omega$ and the gauge fields $W^\Omega$
\be
\epsilon=\epsilon (\varepsilon, W)\,.
\ee
Though $\Ll$ is not gauge invariant under the gauge transformations
of the system, it transforms by a total derivative of a function of
fields $W^\Omega$ and gauge parameters $\varepsilon^\Omega$. As
a result, the action $S$ is gauge invariant.

Note that though the system $\dr w +L(W)=0$ is formally consistent
it is not guaranteed that it admits a solution with regular $w$. In fact,
the Lagrangial $\Ll$ defined by (\ref{L}) is nontrivial for non-exact
 $L(W)$.

The proposal of \cite{Vasiliev:2015mka} is to
 consider invariants resulting from
{the following extension of the HS unfolded equations}
\be\nn
\label{hss}
\W *{} \W = F(\Bb) + \Ll \,Id\q
\W*{} \Bb = \Bb*{} \W\q \dr \Ll=0 \,,
\ee
where
$\W=\dr +W$ {and} $\Bb$ {are differential forms of all odd and even degrees,
respectively (both in} $dx$ {and} $dZ$).
{An appropriate choice is}
\be\nn
 \label{leq}
 iF (\Bb) = dZ_A dZ^A + \eta \delta^2(dz)  \Bb * k*\kappa +
\bar \eta \delta^2( d{\bar z}) \Bb * \bar k * \bar\kappa +G(\Bb) \delta^4(dZ)
k*\bar k * \kappa  *\bar \kappa  \,.
 \ee
$ G = g+ O(\Bb)\,,$
where $g$ {is the coupling constant.}
 $\Ll(x)$ {are} $x${-dependent
space-time differential forms of positive {even} degrees
since the left-hand side of (\ref{hss}) contains a product of forms of
even degrees.} That  it enters as a coefficient
in front of the unit element $Id$ of the star-product algebra means that $\Ll(x)$
is independent of $Y^A$ and $Z^A$. As a result, application of
the covariant derivative to the right-hand side of (\ref{leq}) gives $\dr \Ll(x)=0$.

It should be stressed that the modification of the system by the ``Lagrangians"
$\Ll$ does not affect the form of all equations except for the single
$Z,Y$-independent equation proportional to $Id$, which just acquires the form
(\ref{L}). The form of the Lagrangian $L(W)$ (\ref{L}) now results from
the perturbative solution of the other equations, \ie nonlinear HS equations.

 The {density relevant
 to the generating functional of correlators in} $AdS_4/CFT_3$ {HS holography
 is a  4-form} $\Ll^4$. The {density relevant to BH entropy is a 2-form} $\Ll^2$
 (for recent progress in this direction see \cite{Didenko:2015pjo}.)

 \section{Conclusions}

{A very general property illustrated by the analysis of HS theory is that
the unfolding machinery makes holographic duality manifest at the level of
the unfolded formulation of HS equations. Following
\cite{Vasiliev:2015mka}}, the duality extends to the level
of  generating functionals. The latter can be identified with
integrals of differential
forms of positive even degrees valued in the center of the star-product algebra.
{So defined functionals are {gauge invariant,} coordinate
independent and can be evaluated for any boundaries and bulk solutions}.

In $4d$ HS theory the 4-form $\Ll^4$ is conjectured to give rise to
the generating functional for boundary correlators while the 2-form $\Ll^2$
gives black-hole charges  opening new perspectives for the
understanding of black-hole physics including the informational paradox
\cite{Didenko:2015pjo}.

As shown in \cite{Vasiliev:2015mka} a similar construction applies to the
HS theory in $AdS_3$. In this case the only Lagrangian density is a 2-form
$\Ll^2$. An exciting peculiarity of this construction is that the boundary
functional results from the integration over a one-dimensional cycle at the boundary
(times a cycle over the complexified Poincar\'e coordinate $\zz$). It is tempting to
speculate that this property expresses holomorphicity of the $2d$ boundary
conformal theory.

By virtue of unfolded dynamics usual field-current
correspondence can be extended via interpretation of further nonlinear
combinations of fields with linear fields in higher dimensions. An
interesting subtlety here is that the mixing of fields in different
dimensions representing nonlinear interactions in the original system
can decrease the symmetries of unmixed fields. This is illustrated by
current interactions of massless fields
of all spins in $d=4$ {which break} the $sp(8)$ symmetry of free fields down
{to the conformal symmetry} $su(2,2)\subset sp(8)$ \cite{Gelfond:2015poa}.

\section*{Acknowledgements}
I am grateful to the Institute for Advanced Study of Nanyang
Technical University for its kind hospitality during the workshop
``Higher Spin Gauge Theories".
This research was supported in part by the RFBR Grant No 14-02-01172.


\begin{thebibliography}{99}
\parindent=0pt
\parskip=0pt

\bibitem{Dirac:1936tg}
P.~A.~M.~Dirac,
Proc.\ Roy.\ Soc.\ Lond.\  {\bf 155A} (1936) 447.

\bibitem{Fierz:1939ix}
M.~Fierz and W.~Pauli,
{\em Proc. Roy. Soc. Lond.} {\bfseries A173} (1939) 211--232.


\bibitem{42} V.L.~Ginzburg, JETPh, {\bf 12} (1942) 425.

\bibitem{GT} V.L.~Ginzburg and I.E.~Tamm,  J.Phys. {\bf 11} (1947).

\bibitem{fr} E.S.~Fradkin, JETPh, {\bf 20} (1950) 27; 211.

\bibitem{Rarita:1941mf}
W.~Rarita and J.~Schwinger,
\href{http://dx.doi.org/10.1103/PhysRev.60.61}{{\em Phys. Rev.} {\bfseries 60}
  (1941) 61}.

\bibitem{Fronsdal:1978rb}
C.~Fronsdal, 
{{\em Phys. Rev.} {\bfseries
  D18} (1978) 3624}.

\bibitem{Bengtsson:1983pd}
A.~K.~H. Bengtsson, I.~Bengtsson, and L.~Brink,
{{\em Nucl. Phys.}
  {\bfseries B227} (1983) 31}.

\bibitem{Bengtsson:1983pg}
A.~K.~H. Bengtsson, I.~Bengtsson, and L.~Brink,
{{\em Nucl. Phys.}
  {\bfseries B227} (1983) 41}.

\bibitem{Berends:1984wp}
F.~A. Berends, G.~J.~H. Burgers, and H.~Van~Dam,
{{\em Z. Phys.} {\bfseries C24}
  (1984) 247--254}.

\bibitem{Berends:1984rq}
F.~A. Berends, G.~J.~H. Burgers, and H.~van Dam, 
{{\em Nucl. Phys.}
  {\bfseries B260} (1985) 295}.

\bibitem{Fradkin:1987ks}
E.~S. Fradkin and M.~A. Vasiliev, 
{{\em Phys. Lett.}
  {\bfseries B189} (1987) 89--95}.

\bibitem{Fradkin:1986qy}
E.~S. Fradkin and M.~A. Vasiliev, 
{{\em Nucl. Phys.}
  {\bfseries B291} (1987) 141}.

\bibitem{Coleman:1967ad}
S.~R. Coleman and J.~Mandula,
\href{http://dx.doi.org/10.1103/PhysRev.159.1251}{{\em Phys. Rev.} {\bfseries
  159} (1967) 1251--1256}.


\bibitem{Aragone:1979hx}
C.~Aragone and S.~Deser,
\href{http://dx.doi.org/10.1016/0370-2693(79)90808-6}{{\em Phys. Lett.}
  {\bfseries B86} (1979) 161}.

\bibitem{Maldacena:1997re}
J.~M.~Maldacena,
  Adv.\ Theor.\ Math.\ Phys.\  {\bf 2} (1998) 231
  [Int.\ J.\ Theor.\ Phys.\  {\bf 38} (1999) 1113]
  [arXiv:hep-th/9711200].
\bibitem{Gubser:1998bc}
S.~S.~Gubser, I.~R.~Klebanov and A.~M.~Polyakov,
  Phys.\ Lett.\  B {\bf 428}, 105 (1998)
  [arXiv:hep-th/9802109].

\bibitem{Witten:1998qj}
E.~Witten,
  Adv.\ Theor.\ Math.\ Phys.\  {\bf 2}, 253 (1998)
  [arXiv:hep-th/9802150].

\bibitem{Sundborg:2000wp}
  B.~Sundborg,
  Nucl.\ Phys.\ Proc.\ Suppl.\  {\bf 102} (2001) 113
  [arXiv:hep-th/0103247].

\bibitem{WJ}
E. Witten, talk at the John Schwarz 60-th birthday symposium,
http://theory.caltech.edu/jhs60/witten/1.html



\bibitem{Sezgin:2002rt}
  E.~Sezgin and P.~Sundell,
  Nucl.\ Phys.\  B {\bf 644} (2002) 303
  [Erratum-ibid.\  B {\bf 660} (2003) 403]
  [arXiv:hep-th/0205131].

\bibitem{Klebanov:2002ja}
  I.~R.~Klebanov and A.~M.~Polyakov,
  Phys.\ Lett.\  B {\bf 550} (2002) 213
  [arXiv:hep-th/0210114].

\bibitem{Giombi:2009wh}
  S.~Giombi and X.~Yin,
  JHEP {\bf 1009} (2010) 115
  [arXiv:0912.3462 [hep-th]].

\bibitem{Maldacena:2012sf}
  J.~Maldacena and A.~Zhiboedov,
  Class.\ Quant.\ Grav.\  {\bf 30} (2013) 104003
  [arXiv:1204.3882 [hep-th]].


\bibitem{Vasiliev:2012vf}
  M.~A.~Vasiliev,
  J.\ Phys.\ A {\bf 46} (2013) 214013
  [arXiv:1203.5554 [hep-th]].


\bibitem{Giombi:2012ms}
  S.~Giombi and X.~Yin,
  J.\ Phys.\ A {\bf 46} (2013) 214003
  [arXiv:1208.4036 [hep-th]].

\bibitem{Colombo:2012jx}
  N.~Colombo and P.~Sundell,
  arXiv:1208.3880 [hep-th].

\bibitem{Didenko:2012tv}
  V.~E.~Didenko and E.~D.~Skvortsov,
  arXiv:1210.7963 [hep-th].


\bibitem{Jevicki:2012fh}
  A.~Jevicki, K.~Jin and Q.~Ye,
  J.\ Phys.\ A {\bf 46} (2013) 214005
  [arXiv:1212.5215 [hep-th]].

\bibitem{Giombi:2013fka}
  S.~Giombi and I.~R.~Klebanov,
  JHEP {\bf 1312} (2013) 068
  [arXiv:1308.2337 [hep-th]].

\bibitem{Giombi:2014yra}
  S.~Giombi, I.~R.~Klebanov and A.~A.~Tseytlin,
  Phys.\ Rev.\ D {\bf 90} (2014) 024048
  [arXiv:1402.5396 [hep-th]].

\bibitem{Beccaria:2014jxa}
  M.~Beccaria, X.~Bekaert and A.~A.~Tseytlin,
  JHEP {\bf 1408} (2014) 113
  [arXiv:1406.3542 [hep-th]].


\bibitem{Koch:2014aqa}
  R.~d.~M.~Koch, A.~Jevicki, J.~P.~Rodrigues and J.~Yoon,
  arXiv:1408.4800 [hep-th].

\bibitem{Giombi:2014xxa}
  S.~Giombi and I.~R.~Klebanov,
  arXiv:1409.1937 [hep-th].

\bibitem{Beccaria:2014xda}
  M.~Beccaria and A.~A.~Tseytlin,
  arXiv:1410.3273 [hep-th].

\bibitem{Barvinsky:2014kta}
  A.~O.~Barvinsky,
  J.\ Exp.\ Theor.\ Phys.\  {\bf 120} (2015) 3,  449
  [arXiv:1410.6316 [hep-th]].

\bibitem{Bekaert:2015tva}
  X.~Bekaert, J.~Erdmenger, D.~Ponomarev and C.~Sleight,
  JHEP {\bf 1511} (2015) 149
  [arXiv:1508.04292 [hep-th]].

\bibitem{Hegde:2015dqh}
  A.~Hegde, P.~Kraus and E.~Perlmutter,
  JHEP {\bf 1601} (2016) 176
  [arXiv:1511.05555 [hep-th]].


\bibitem{Henneaux:2010xg}
  M.~Henneaux and S.~J.~Rey,
  JHEP {\bf 1012} (2010) 007
  [arXiv:1008.4579 [hep-th]].

\bibitem{Campoleoni:2010zq}
  A.~Campoleoni, S.~Fredenhagen, S.~Pfenninger and S.~Theisen,
  JHEP {\bf 1011} (2010) 007
  [arXiv:1008.4744 [hep-th]].


\bibitem{Gaberdiel:2010pz}
  M.~R.~Gaberdiel and R.~Gopakumar,
  Phys.\ Rev.\  D {\bf 83} (2011) 066007
  [arXiv:1011.2986 [hep-th]].

\bibitem{Koh:1986vg}
  I.~G.~Koh and S.~Ouvry,
  Phys.\ Lett.\ B {\bf 179} (1986) 115
   [Phys.\ Lett.\  {\bf 183B} (1987) 434].

\bibitem{Bengtsson:1987jt}
  A.~K.~H.~Bengtsson,
  Class.\ Quant.\ Grav.\  {\bf 5} (1988) 437.

\bibitem{Metsaev:2005ar}
  R.~R.~Metsaev,
  Nucl.\ Phys.\  B {\bf 759} (2006) 147
  [arXiv:hep-th/0512342].

\bibitem{Sagnotti:2010at}
  A.~Sagnotti and M.~Taronna,
  Nucl.\ Phys.\  B {\bf 842} (2011) 299
  [arXiv:1006.5242 [hep-th]].

\bibitem{Fotopoulos:2010ay}
  A.~Fotopoulos and M.~Tsulaia,
  JHEP {\bf 1011} (2010) 086
  [arXiv:1009.0727 [hep-th]].

\bibitem{Manvelyan:2010je}
  R.~Manvelyan, K.~Mkrtchyan and W.~Ruehl,
  Phys.\ Lett.\  B {\bf 696} (2011) 410
  [arXiv:1009.1054 [hep-th]].

\bibitem{Boulanger:2011dd}
  N.~Boulanger and P.~Sundell,
  J.\ Phys.\ A  {\bf 44} (2011) 495402
  [arXiv:1102.2219 [hep-th]].

\bibitem{Sezgin:2011hq}
  E.~Sezgin and P.~Sundell,
  arXiv:1103.2360 [hep-th].

\bibitem{Vasilev:2011xf}
  M.~A.~Vasiliev,
  Nucl.\ Phys.\ B {\bf 862} (2012) 341
  [arXiv:1108.5921 [hep-th]].

\bibitem{Joung:2012hz}
  E.~Joung, L.~Lopez and M.~Taronna,
  JHEP {\bf 1301} (2013) 168
  [arXiv:1211.5912 [hep-th]]


\bibitem{Vasiliev:2015mka}
  M.~A.~Vasiliev,
  arXiv:1504.07289 [hep-th].

\bibitem{more} M.~A.~Vasiliev, {\it Phys. Lett.}  B {\bf 285} (1992) 225.

\bibitem{Ann}
M.~A.~Vasiliev,
Ann. Phys. (NY) {\bf 190} {(1989)} {59}.

\bibitem{Vasiliev:2005zu}
  M.~A.~Vasiliev,
  Int.\ J.\ Geom.\ Meth.\ Mod.\ Phys.\  {\bf 3} (2006) 37
  [hep-th/0504090].

\bibitem{Shaynkman:2001ip}
  O.~V.~Shaynkman and M.~A.~Vasiliev,
  Theor.\ Math.\ Phys.\  {\bf 128} (2001) 1155
   [Teor.\ Mat.\ Fiz.\  {\bf 128} (2001) 378]
  [hep-th/0103208].

\bibitem{Vasiliev:2001zy}
  M.~A.~Vasiliev,
  Phys.\ Rev.\ D {\bf 66} (2002) 066006
  [hep-th/0106149].


\bibitem{Gelfond:2003vh}
  O.~A.~Gelfond and M.~A.~Vasiliev,
  Theor.\ Math.\ Phys.\  {\bf 145} (2005) 1400
   [Teor.\ Mat.\ Fiz.\  {\bf 145} (2005) 35]
  [hep-th/0304020].

\bibitem{FF}M. Flato and C. Fronsdal,
 {\it Lett. Math. Phys.}{\bf 2} (1978) 421.

\bibitem{Fr1} C.~Fronsdal, ``Massless Particles, Ortosymplectic Symmetry
  and Another Type of Kaluza--Klein Theory'', Preprint UCLA/85/TEP/10,
  in Essays on Supersymmetry, Reidel, 1986 (Mathematical Physics
  Studies, v.8).


\bibitem{BL}I.~Bandos and  J.~Lukierski,
{\it Mod.Phys. Lett} {\bf A14} (1999) 1257,
{ hep-th/9811022}.

\bibitem{BLS}I.~Bandos, J.~Lukierski and D.~Sorokin,
{\it Phys. Rev.} {\bf D61} (2000) 045002, { hep-th/9904109}.



 \bibitem{Mar}
M.A. Vasiliev,`` Relativity, Causality, Locality, Quantization and
Duality in the $Sp(2M)$ Invariant Generalized Space-Time'', {
hep-th/0111119}.


\bibitem{Gelfond:2010pm}
  O.~A.~Gelfond and M.~A.~Vasiliev,
  J.\ Exp.\ Theor.\ Phys.\  {\bf 120} (2015) 3,  484
  [arXiv:1012.3143 [hep-th]].

\bibitem{Gelfond:2015poa}
  O.~A.~Gelfond and M.~A.~Vasiliev,
  arXiv:1510.03488 [hep-th].

\bibitem{Sezgin:2003pt}
  E.~Sezgin and P.~Sundell,
  JHEP {\bf 0507} (2005) 044
  [arXiv:hep-th/0305040].

\bibitem{Vasiliev:1999ba}
  M.~A.~Vasiliev,
  In *Shifman, M.A. (ed.): The many faces of the superworld* 533-610
  [hep-th/9910096].

\bibitem{Didenko:2014dwa}
  V.~E.~Didenko and E.~D.~Skvortsov,
  arXiv:1401.2975 [hep-th].

\bibitem{Didenko:2015pjo}
  V.~E.~Didenko, N.~G.~Misuna and M.~A.~Vasiliev,
  arXiv:1512.07626 [hep-th].

\end{thebibliography}
\end{document}